\documentstyle[preprint,aps]{revtex}
\font\elevenbf=cmbx10 scaled\magstep 1
\newcommand{\R}{\mbox{$R$}}
\newcommand{\Rb}{\mbox{$\not \! \! \R$}}
\newcommand{\Lb}{\mbox{$L$}}
\newcommand{\Lbb}{\mbox{$\not \! \! \Lb$}}
\newcommand{\B}{\mbox{$B$}}
\newcommand{\Bb}{\mbox{$\not \! \! \B$}}

\newcommand{\be}{\begin{equation}}
\newcommand{\ene}{\end{equation}}
\newcommand{\een}{\end{subequations}}
\newcommand{\ben}{\begin{subequations}}
\newcommand{\beq}{\begin{eqnarray}}
\newcommand{\eeq}{\end{eqnarray}}

\begin{document}
\preprint{PRL-TH-94/11,hep-ph/9403349}
\title
{\bf LEPTONIC CP-VIOLATION IN SUPERSYMMETRIC  STANDARD MODEL\\}
\author{ Anjan S. Joshipura and  Marek Nowakowski}
\address { Theory Group, Physical Research Laboratory,
Navrangpura, Ahmedabad 380 009, India}
\maketitle
\begin{abstract}
We point out the possibility of spontaneous and hard CP-violation
in the scalar potential of R-parity broken supersymmetric
Standard Model. The existence of spontaneous CP-violation depends
crucially on the R-parity breaking terms in the
superpotential and, in addition, on the choice of the soft supersymmetry
breaking terms. Unlike in theories with R-parity conservation,
it is natural, in the context of the present model, for the
sneutrinos to acquire (complex) vacuum expectation values. In
the context of this model we
examine here the global implications, like the strength of the
CP-violating interactions and the neutrino masses.
\end{abstract}
\newpage
The minimal standard electroweak model provides adequate description of
CP-violation hitherto seen in laboratory \cite{paschos}. In addition to this
CP-violation,
there seems to be a good reason to expect CP-violation in the
leptonic sector. The motivation for having such CP-violation
comes from the desire to generate the observed baryon asymmetry
in the universe at electroweak scale \cite{cohen}. It is well known that the
sphaleron induced baryon number violation tends to erase the
baryon asymmetry generated at the GUT scale in theories with exact
$B-L$ symmetry. But if lepton number violating interactions
generate some lepton asymmetry, this could be transformed to
baryon asymmetry by sphaleron induced reactions. Generation of
lepton number asymmetry needs both, lepton non-conserving
interactions and CP- as well as C-violation in them. Neither of
these are present in the Standard Model. Thus it is important to
look for models which contain both, L- and CP-violating interactions. Such a
study is important in its own right, independent of the
arguments given above, since it might be easier to detect
additional CP-violating processes in the leptonic sector. In
this note we study the nature of CP-violation in lepton
non-conserving and R-parity violating ($\Rb$) Minimal Supersymmetric
Standard Model (MSSM).

The CP-violation in the MSSM has been extensively discussed in
the literature. The presence of supersymmetric particles leads
to new sources of CP-violation \cite{susycp}. These have been shown to be
insufficient for the explanation of CP-violation in
$K^0-\overline{K^0}$ system \cite{susycp}. On the other hand they lead to
large electric dipole moments (edm) for the neutron and electron
\cite{susyedm}. All
the discussions in the literature are confined to R-parity
conserving MSSM. Introduction of R-violating terms automatically
generates lepton ($\Lbb$) or baryon ($\Bb$) number violating interactions.
These
new interactions change the features of CP-violation in the MSSM
in a qualitative manner. They introduce additional parameters
which allow both, explicit CP-violation in the Higgs potential
as well as the possibility of breaking CP spontaneously. We
study here this CP-violation as well as constraints on its magnitude.
Since this kind of CP-violation is exclusively connected to
lepton number violating interactions its effects will show up
only in the context of $\Lbb$-reactions. Hence its effect on
`usual' CP-violation is negligible.

R-parity assigns the quantum number $+1$ to conventional
particles and $-1$ to their superpartners. More specifically it can
be written as
\be \label{e0}
R=(-1)^{3B+L+2S} \ene
where $B$, $L$ and are the baryon and lepton number and $S$ is
the spin.
Let us then split the superpotential $W$ of the MSSM into a R-parity
conserving part ($W_0$) and R-parity violating term ($W_{\Rb}$) i.e.
\be \label{e1}
W=W_0+W_{\Rb} \ene
In the following we use a symbol with a hat, $\hat{A}$, to
indicate a chiral superfield and the same symbol without a hat,
$A$, for the spin-zero field content of the chiral
supermultiplet. Let then $\hat{L}_i$ ($\hat{E}_i^C$) and
$\hat{Q}_i$ ($\hat{U}_i^C $,$ \hat{D}_i^C$) denote the lepton and
quark doublets (lepton and quarks $SU(2)$ singlets) with
generation index $i$, respectively and let $\hat{H}_{1,\; 2}$ be
the super-Higgs fields. The standard form for $W_0$ is
\be \label{e2}
W_0 = \epsilon_{ab}\left [h_{ij}\hat{L}_i^a \hat{H}_1^b
\hat{E}_j^C + h'_{ij}\hat{Q}_i^a \hat{H}_1^b \hat{D}_j^C + h''_{ij}
\hat{Q}_i^a \hat{H}_2 \hat{U}_j^C + \mu \hat{H}_1^a \hat{H}_2^b \right]
\ene
where $a$, $b$ are $SU(2)$ group indices. The $U(1)_Y$ quantum
number assignment is as usual: $Y(\hat{L}_i)=-1$,
$Y(\hat{E}_i^C)=2$, $Y(\hat{Q}_i)=1/3$, $Y(\hat{D}_i^C)=2/3$,
$Y(\hat{U}_i^C)=-4/3$, $Y(\hat{H}_1)=-1$, $Y(\hat{H}_2)=1$.

In general, the R-parity violating part $W_{\Rb}$ reads \cite{weinberg1}
\be \label{e3}
W_{\Rb}=\epsilon_{ab}\left[\lambda_{ijk}\hat{L}_i^a \hat{L}_j^b
\hat{E}_k^C + \lambda'_{ijk}\hat{L}_i^a \hat{Q}_j^b \hat{D}_k^C
+ \mu_i \hat{L}_i^b \hat{H}_2^b \right] + \lambda''_{ijk}
\hat{U}_i^C \hat{D}_j^C \hat{D}_k^C \ene
The terms in eq. (\ref{e3}) proportional to
$\lambda_{ijk}=-\lambda_{jik}$, $\lambda'_{ijk}$ and $\mu_i$  violate
lepton number whereas the baryon number is explicitly broken by the
$\lambda''_{ijk}$-term ($\lambda''_{ijk}=-\lambda''_{ikj}$).
It is well known that keeping both these
terms in the lagrangian (i.e. $\Lbb$ and $\Bb$ interaction terms)
leads to difficulties with proton lifetime \cite{hall}. Therefore we will
set from now on $\lambda''_{ijk}=0$.

The term $\epsilon_{ab} \mu_i \hat{L}_i^a \hat{H}_2^b$ is also
not included conventionally. This is due to the fact that this
term can always be rotated away from the superpotential by
redefinition of the Higgs $\hat{H}_1$ and the leptonic superfields
$\hat{L}_i$. It is worth stressing, however, that such a
redefinition does not leave the full lagrangian (including soft
breaking terms) invariant. Apart from changing $\lambda_{ijk}$,
$\lambda'_{ijk}$ in eq. (\ref{e3}) in a well known way this redefinition
also affects the soft supersymmetry breaking terms which are
usually induced through supergravity. Given the superpotential
in eqs. (\ref{e2}) and (\ref{e3}),
the soft terms involving
the scalars have the following general form in MSSM
\beq \label{e4}
V_{soft}&=&m_1^2 H_1^{\dagger}H_1 + m_2^2 H_2^{\dagger} H_2
+m_{L_i}^2L_i^{\dagger}L_i \nonumber \\
&-&(m_{12}^2\epsilon_{ab}H_1^a H_2^b +h.c.) + (\kappa'_i \epsilon_{ab}
H_2^a L_i^b +h.c.) \nonumber \\
&+& {\rm cubic\;\; terms} \eeq
The parameters $\kappa'_i$ and $m_{12}^2$ would be related to
the parameters of the superpotential (\ref{e2}) and (\ref{e3}) at
Planck scale in the usual way. The cubic terms are
soft breaking terms in correspondence to cubic terms in $W$.
While the $\mu_i$ term in eq. (\ref{e3}) can always be rotated
away, the corresponding $\kappa'_i$-terms would still be present
in the low energy theory.
Removal of the $\epsilon_{ab} \mu_i
\hat{L}_i^a \hat{H}_2^b$ term in eq. (\ref{e3}) needs a redefinition
\be \label{e5}
\mu' \hat{H}'_1=\mu \hat{H}_1 + \mu_i \hat{L}_i \ene
Each of the $\hat{L}_i$ fields have to be replaced by a
combination orthogonal to (\ref{e5}). It is easy to see that this
orthogonal transformation does not leave the soft breaking terms
in (\ref{e5}) invariant. Hence even if one removes the
the $\mu_i$-terms from $W_{\Rb}$ (i.e. sets $\kappa'_i=0$)
the term $\epsilon_{ab} L_i^aH_2^b$ as well as an additional
term $L_i^{\dagger}H_1$ will get generated in $V_{soft}$.
Conversely, if one does not rotate the term proportional to
$\mu_i$ in (\ref{e3}), the $L_i^{\dagger}H_1$ part will arise from
the F-term associated with $H_2$ and the $\kappa'_i$ term would
come from the general soft breaking expressions. In either case,
one would obtain two additional complex parameters. We prefer to
retain the $\epsilon_{ab} \hat{L}_i^a \hat{H}_1^b$ in (\ref{e3})
and discuss its implications. The effect of these additional
terms in $V_{soft}$ generated in the process of removing the
$\mu_i$-dependent terms in $W_{\Rb}$ is not investigated in the
literature (see however \cite{masiero}). These terms play an
important role in generating spontaneous CP-violation as we will
see.

With the change of notation $\varphi_i \equiv L_i$, $\phi_2
\equiv H_2$ and $\phi_1 \equiv -i\tau_2 H_1^*$ ($\tau_2$ being
the Pauli matrix, $(i\tau_2)_{ab}=\epsilon_{ab}$) we derive the
Higgs potential
\beq \label{e6}
V_{Higgs}&=&\mu_1^2\vert \phi_1 \vert^2 + \mu_2^2 \vert \phi_2 \vert^2
+\mu_{L_i}^2 (\varphi_i^{\dagger}\varphi_i) \nonumber \\
&+& {1 \over 2}\lambda_1\left[\vert \phi_1 \vert^4 + \vert \phi_2
\vert^4 + (\varphi_i^{\dagger} \varphi_i)^2 + 2\vert \phi_1 \vert^2
(\varphi_i^{\dagger} \varphi_i) -2\vert \phi_2 \vert^2(\varphi_i^{\dagger}
\varphi_i) \right] \nonumber \\
&+&\lambda_2 \vert \phi_1 \vert^2 \vert \phi_2 \vert^2 -(\lambda_1
+\lambda_2)\vert \phi_1^{\dagger} \phi_2 \vert^2 +\left (\lambda_3
(\phi_1^{\dagger} \phi_2) + h.c. \right) \nonumber \\
&+&\left(i\kappa_i(\phi_1^T\tau_2 \varphi_i) +h.c. \right)
+ \left(i\kappa'_i(\phi_2^T \tau_2 \varphi_i) +h.c. \right)+V_{rest} \eeq
In the above equation it is assumed that all $\mu_i$ are the
same for all generations, $\mu_i \equiv \mu_0$. This is not essential
and we have done it to simplify things \cite{note1}. The parameters in
(\ref{e6}) like $\lambda_i$ ($i=1,2,3$), $\mu_i$ ($i=1,2$),
$\mu_{L_j}$ and $\kappa'_j$ ($j$ is the generation index) can be
expressed as in the standard case in terms of the $SU(2)$
($U(1)_Y$) coupling constant $g$ ($g'$) as well as the
parameters entering eqs. (\ref{e2})-(\ref{e4}).
\beq \label{e7}
&&\mu_1^2=m_1^2+\vert \mu \vert^2,\; \; \; \mu_2^2=m_2^2 +\vert
\mu \vert^2 + \mu_i \mu_i^*, \; \; \; \mu_{L_i}^2=m_{L_i}^2+
\vert \mu_i \vert^2 \nonumber \\
&&\lambda_1={1 \over 4}(g^2 + g'^2), \; \; \; \lambda_2={1 \over
2}g^2- \lambda_1,\; \; \;  \lambda_3=-m_{12}^2, \; \; \;
\kappa_i=\mu^* \mu_i \eeq
$V_{rest}$ in (\ref{e6}) contains all terms of the potential which
are not relevant for minimization. The full form of $V_{rest}$
will be given elsewhere. Here we merely write two terms to
display the presence of hard CP-violation in $V_{Higgs}$
\be \label{e8}
V_{rest}=\kappa_{ij}(\phi_1^{\dagger}\varphi_j)(\varphi_i^{\dagger}
\phi_1) +\kappa_{nmij}(\varphi_i^T \tau_2 \varphi_j)(\varphi_n^T \tau_2
\varphi_m)^{\dagger} + ... \ene
with
\beq \label{e9}
&&\kappa_{jk}=\kappa_{kj}^* \equiv h_{ji}^* h_{ki} \nonumber \\
&&\kappa_{nmij}=-\kappa_{mnij}=-\kappa_{nmji}=\kappa_{ijnm}^* \equiv
\lambda_{nmk}^* \lambda_{ijk} \eeq
where $h_{ij}$ is the leptonic Yukawa coupling in (\ref{e2}) and
$\lambda_{ijk}$ enters  eq.(\ref{e3}). Note that the leptonic
Yukawa coupling need not be diagonal.

The potential in (\ref{e6}) contains two additional (in general complex)
parameters $\kappa_i$ and $\kappa'_i$ for every generation index
$i$. Their presence gives rise to three important features not present
in the R-conserving MSSM. ({\em i}) Firstly, when both $\kappa_i$ and
$\kappa'_i$
are present, $V_{Higgs}$ is not invariant under CP. ({\em ii}) Even if
CP is imposed on $V_{Higgs}$, the simultaneous presence of
$\kappa_i$ and $\kappa'_i$ allows the possibility of spontaneous
CP-violation. ({\em iii}) These new terms are linear in the sneutrino
fields and as consequence $\varphi_i$ acquire vacuum expectation
values (vev). This in turn generates masses for neutrinos via neutralino
neutrino mixing. We discuss this features in what follows.

It is easy to see that $V_{Higggs}$ (in the first step without
$V_{rest}$) is CP-invariant only if
\be \label{e10}
\Im  m(\kappa_i \kappa'_j \lambda_3)\delta_{ij}=0 \ene
Hence, when both $\kappa_i$ and $\kappa'_i$ are present the
potential violates CP. Independently of (\ref{e10})
$V_{rest}$ is CP-invariant only if the following condition holds
\be \label{e11}
\Im  m (\kappa_{n'm'i'i'}\kappa^*_{ni}\kappa^*_{mj})\delta_{n'n}
\delta_{m'm}\delta_{i'i}\delta_{j'j}=0 \ene
Other similar conditions can be derived which signal the
presence of
hard CP-violation in the potential (i.e. CP-violation which is independent
of the possibility of spontaneous CP-violation). In order to
demonstrate the latter, let us assume the parameters $\kappa_i$, $\kappa'_i$
and $\lambda_3$ to be real. Let us denote the vacuum expectation
values (vev's) of the fields by $<\phi^T_{1,\; 2}>=(0,\; \;
v_{1,\; 2})$ and $<\varphi^T_i>=(w_i,\;\; 0)$. Then at the minimum
\beq \label{e12}
&v_1^*&\left[\mu_1^2+\lambda_1(\vert v_1 \vert^2 -\vert v_2 \vert^2
+w_i^* w_i )\right]+\lambda_3 v_2^* - \kappa_i w_i =0 \nonumber \\
&v_2^*&\left[\mu_2^2 - \lambda_1 (\vert v_1 \vert^2 - \vert v_2 \vert^2
+w_j^* w_j )\right] + \lambda_3 v_1^* - \kappa'_j w_j= 0 \nonumber\\
&w_j^*&\delta_{ij} \left[\mu_{L_i}^2 + \lambda_1(\vert v_1 \vert^2
-\vert v_2 \vert^2 + w_k^* w_k)\right] -\kappa_i v_1 -\kappa'_iv_2=0
\eeq
It follows from these conditions that the $w_i$'s are automatically
non-zero as long as $\kappa_i$'s, $\kappa'_i$'s and $v_{1,\; 2}$
are non-zero. Setting $w_i$ to zero leads to adjusting
the parameters of the potential and hence one must allow vev's
for all three sneutrino fields. It follows namely from (\ref{e12})
by setting $w_i=0$ that (without loss of generality for real parameters)
\be \label{e13}
(\mu_1^2 + \mu_2^2)\kappa_i \kappa'_j \delta_{ij}=\lambda_3(\kappa_i^2+
\kappa^{\prime 2}_i) \ene
In other words it is natural in the context of the potential
(\ref{e6}) for the sneutrinos to acquire vev's. This situation
should be contrasted
with MSSM without $\Rb$ i.e. putting
$\kappa_i=\kappa'_i=0$ in eq. (\ref{e12}).
Even there it is possible to obtain a non-zero vev $w_i$ provided
the following equation is satisfied \cite{masiero}
\be \label{e14}
(\mu_1^2 - \mu_{L_k}^2)(\mu_2^2 + \mu_{L_j}^2)=\lambda^2_3 \ene
for any two generation indices $i$, $j$. Here $w_i=0$ would be
the natural choice \cite{note2}. If $w_i \neq 0$ then this vev
is expected to be large \cite{ross}, in general, and would conflict with
phenomenology (see later) unless parameters are restricted
\cite{masiero}. We
shall assume that the parameters satisfy such restriction derived
in ref. \cite{masiero} and that $w_i=0$ when $\kappa_i$ and $\kappa'_i$
are zero. In such a situation, the lepton number is not
spontaneously broken and the spectrum does not contain any
majoron. This has important phenomenological implications which
we will discuss later.

There is yet another, physical motivation why sneutrinos should have
non-zero complex vev's once R-parity is explicitly broken in the
lagrangian. Dropping the crucial term $\epsilon_{ab}\hat{L}_i^a
\hat{H}_2^b$, but retaining the $\lambda_{ijk}$ and $\lambda'_{ijk}$
terms in eq. (\ref{e3}) and assuming $w_i =0$ the relevant sneutrino
mass terms are simply
\be \label{ne1}
\left[\mu_{L_i}^2 + \lambda_1 (v_1^2-v_2^2)\right](\varphi_{0i}^{R}
\;\varphi_{0i}^{R} + \varphi_{0i}^{I}\;\varphi_{0i}^{I}) \ene
where $\varphi_{0i}^R$ and $\varphi_{0i}^I$ are the real and
imaginary parts of the neutral component of $\varphi_i$, respectively.
They correspond to states with
$L=1$ and $L=-1$ quantum numbers. We see from eq. (\ref{ne1}) that in spite
of having $\Lbb$-terms in the lagrangian these states do not
mix. Since such mixing would be natural in lepton number
violating theory
we need complex vev's of sneutrinos (note that
for instance
the terms proportional $\kappa_i$ mix real and imaginary
components of $\phi_1$ and $\varphi_i$).

Indeed eqs. (\ref{e12}) allow for complex vev's as long as the
crucial parameters $\kappa_i$ and $\kappa'_i$ are non-zero. To see
this explicitly set $v_1$ real, $v_2=\vert v_2 \vert
e^{i\alpha}$, $w_3 \equiv w=\vert w \vert e^{i \gamma}$, $w_{1,\;
2}=0$, $\kappa_{1,\; 2}=\kappa'_{1,\; 2}=0$ and $\kappa_3\equiv \kappa$
and $\kappa'_3 \equiv \kappa'$ (say, in one
generation case). Then from (\ref{e12}) we get
\beq \label{e15}
&&\vert v_2 \vert \lambda_3 \sin \alpha + \vert w \vert \kappa \sin
\gamma =0 \nonumber \\
&&\vert v_1 \vert \lambda_3 \sin \alpha - \vert w \vert \kappa' \sin
(\alpha + \gamma)=0 \eeq
Solving this for the phases one obtains
\beq \label{e16}
&&\cos \alpha={A^2(B^2-1)+1 \over 2AB} \nonumber \\
&&\cos \gamma={1 \over A}{A^2(B^2+1)+1 \over 2AB} \eeq
where $A$ and $B$ are defined through
\be \label{e17}
A \equiv -{\vert w \vert \kappa \over \vert v_2 \vert \lambda_3},
\;\;\;\; B\equiv {v_1 \lambda_3 \over \kappa' \vert w \vert} \ene

In general, the amount of CP-violation (eq. (\ref{e16})) characterized
through $\kappa$, $\kappa'$ and $w$ is restricted from phenomenology.
The restriction on sneutrino vev come from ({\em a}) LEP data
around the the $Z^0$ resonance and \cite{ellis1} ({\em b }) from restrictions
on neutrino masses \cite{barbieri,ellis2}. If $w \neq 0$ when
$\kappa=\kappa'=0$ then
the theory contains a majoron. In this case $Z^0$ could decay
into a majoron and an associated scalar. The invisible $Z^0$ width
strongly constrains this possibility. Even if there is no
majoron, as in the present case, the LEP data do imply
significant restrictions \cite{barbieri}. However, more important restrictions
come from the neutrino masses. The presence of the parameter
$\mu_i$ in $W_{\Rb}$ leads directly to mixing between neutrino
and higgsino and as a consequence, to neutrino masses. In addition,
$\kappa_i$ and $\kappa'_i$ induce, as shown before, sneutrino vev's
which mix neutrinos with gauginos $\tilde{\lambda}_a$. Thus neutrino
masses constrain the parameters $\kappa_i/\mu$ and $w_i$. We
assume only one generation for simplicity.
Then the neutralino mass matrix in one generation case
and the $(\tilde{B},\;\tilde{W}_3,\;\tilde{H}_1^0,\;\tilde{H}_2^0,\;
\nu)$ basis takes the following form
\be \label{e18}
M_{\tilde{\lambda}_a / \nu}=\left(\begin{array}{ccccc}
cM & 0 & -g'v_1/2 & g'v_2/2 & -g'w/2\\
0 & M & gv_1/2 & -gv_2/2 & gw/2 \\
-g'v_1/2 & gv_1/2 & 0 & -\mu & -\kappa /\mu \\
g'v_2 & -gv_2/2 & -\mu & 0 & 0 \\
-g'w/2 & gw/2 & -\kappa /\mu & 0 & 0 \end{array}\right ) \ene
where we have dropped all possible CP-violating phases.
This has been analyzed in ref. \cite{barbieri} in the limit
$\kappa \to 0$. The presence of $\kappa$ makes a minor modification.
The parameter $c$ has been taken in \cite{barbieri} to be $0.49$ with the
assumption that the gaugino masses scale like gauge couplings.
The
mass matrix (\ref{e18}) leads to the following neutrino mass
\be \label{e19}
m_{\nu} \simeq {g^2 \over 4 \mu^2 \cos^2 \theta_W}\;\; {(\mu w+
\kappa v_1 /\mu)^2 \over \vert (M_Z^2/\mu)\sin 2\beta - bM \vert} \ene
with
\be \label{e20}
b={c \over c \cos^2 \theta_W + \sin^2 \theta_W},\; \; \; \tan
\beta={v_1 \over v_2} \ene
The neutrino masses are required to be $\le {\cal O}(10 {\rm
eV})$. Otherwise they will overclose the universe. For
$M \sim \mu \sim {\rm TeV}$ and $\tan \beta =1$ dominant
contribution to $m_{\nu}$ comes from the $w$ term in eq. (\ref{e19})
and one obtains
\be \label{e21}
w \le {\cal O}({\rm MeV}) \ene
We note here that due to absence of the majoron, the decay of a
heavier neutrino into a lighter one plus majoron is not
possible in the present case. But the presence of flavor changing couplings
of neutrinos to $Z^0$ \cite{ellis2} may allow
fast decay into three neutrinos. If this happens then the limit on
$w$ could be relaxed.

Much stronger constraints on $\kappa_i$ and $\kappa'_i$ can be
derived from considerations based on baryon asymmetry \cite{masiero},
\cite{campbell}.
The lepton number violating interaction as well as sphaleron induced
$B+L$ violating processes, if simultaneously in equilibrium,
will wash out the original baryon asymmetry. Demanding that the processes
induced by the $\mu_i$ terms in eq. (\ref{e3}) be out of
equilibrium typically requires \cite{masiero}
\be \label{e22}
\kappa_i \le 10^{-6}{\rm MeV}^{2} \ene
Similar constraint holds for $\kappa'_i$. However, these are
model independent constraints. If there is some unbroken global
symmetry associated with family lepton number (e.g.
$\kappa_1=0$) then the constraint (\ref{e22}) does not apply \cite{nelson}.
But constraint (\ref{e21}) still holds. Assuming $w \sim \mu_i
\sim \kappa /\mu$ one sees from eq. (\ref{e17}) that $A \sim 1/B
\le \displaystyle{\left({{\rm MeV} \over {\rm TeV}}\right)^2} \simeq 10^{-12}$.
Hence it follows from eqs. (\ref{e15},\ref{e16}) that the phase
$\alpha$ between the vev's of $\phi_1$ and $\phi_2$ is extremely
small.
This phase would be associated with CP-violation in lepton number
conserving processes. In contrast, the relative phase $\gamma$
is ${\cal O}(AB)$ and could therefore be large. But this phase will
invariably be accompanied by
lepton number violation signified by the sneutrino vev. {\em
Hence one would expect the CP-violation in $\Lbb$-processes to
be large}.

At this point it might be instructive to compare other efforts to
introduce spontaneous CP-violation in MSSM. First note that in a
general two Higgs doublet model with softly broken $Z_2$
symmetry \cite{weinberg2} the CP-violation comes from the following
two terms of the potential
\be \label{e23}
{1 \over 2}\lambda_5\;(\phi_1^{\dagger}\phi_2)^2 + \lambda_6\;(\phi_1^{\dagger}
\phi_2)\;\; + \;\; h.c. \ene
Im MSSM without R-parity breaking $\lambda_6=\lambda_3$ form eq.
(\ref{e6}), but $\lambda_5=0$ at tree level. The idea is then to
generate this term radiatively \cite{maekawa}. For real
$\lambda_5$ and $\lambda_6$ spontaneous CP-violation is possible
modulo a restriction on the parameters which essentially comes
from the obvious inequality $\vert \cos \xi \vert \le 1$ where
$\xi$ is the CP-violating phase. It then turns out that $\lambda_5$
is very small which together with the parameter constraint leads
to a very light Higgs boson inconsistent with LEP data
\cite{pomarol1}. On the other hand, one can induce radiatively other
(complex) couplings like $\lambda_7$ and $\lambda_8$ giving rise
to the interaction terms of the form
\be \label{e24}
\lambda_7\; (\phi_1^{\dagger}\phi_2)(\phi_1^{\dagger}\phi_1) +
\lambda_8\; (\phi_1^{\dagger}\phi_2)(\phi_2^{\dagger}\phi_2) + h.c.\ene
This is possible since the CP is violated in other sectors of
the MSSM lagrangian. The amount of such CP-violation is then heavily
constrained by limits of edm of the neutron and turns out to be too
small to be of any significance for phenomenology \cite{pomarol1}.
Note that the model discussed here evades the limits coming from edm in a
natural way.

Enlarging the superpotential by interaction terms with a singlet
field $\hat{N}$ it is possible to have spontaneous CP-violation
\cite{pomarol2}. These new interaction terms are, in general, linear
combination of the following invariants
\beq \label{e25}
&\hat{N}&\hat{H}_1 \tau_2 \hat{H}_2, \nonumber \\
&\hat{N}^3&,\;\; \hat{N}^2,\;\; \hat{N} \eeq
Indeed the additional superpotential $W_N$ consisting of the
first three terms in eq. (\ref{e25}) has been shown to give rise
to spontaneous CP-violation \cite{pomarol2}. Furthermore such a model,
with real Kobayashi-Maskawa matrix, can explain the CP-violation
in $K^0-\overline{K^0}$ system. In ref. \cite{ramao} it has been
proved that any combination of the first term in eq.
(\ref{e25}) with one of the other terms (involving only the singlet
field) does not lead to spontaneous CP-violation at tree level.
Higher order corrections to the potential can change this result
\cite{babu} and spontaneous CP-violation becomes possible in
such model. This model requires relatively light Higgs bosons
due to a sum rule $m_{H_1} + m_{H_2} \le 100 {\rm GeV}$ \cite{babu}
One should bear in mind that generating spontaneous CP-violation
through higher order corrections can be delicate matter due to
the Georgi-Pais result. The latter states that provided the loop
corrections are small and the true minimun is close to its tree
level value spontaneous CP-violation cannot be produced through
quantum effects unless a massless particle different from the
Goldstone mode appears in the spectrum. With the present limit
on the top mass one can, however, argue that the loop
corrections to the potential are not small any more \cite{babu}.

Finally we mention that spontaneous CP-violation in the context
of MSSM at finite temperature has been proposed and discussed in
\cite{temperature}

In summary, we have shown that the MSSM contains additional
sources of CP-violation associated with R-parity and lepton
number breaking processes. This CP-violation is argued to be constrained
by neutrino masses and sphaleron induced transitions.
CP-violation associated with $\Lbb$-processes could be large.
Such a situation would arise typically in a $\Rb$ transition
such as the decay of the lightest supersymmetric particle
induced through the type of interaction terms we have considered.
This may have no significance as far as laboratory experiments are
concerned, but it may have cosmological implications for baryo-genesis
via lepto-genesis \cite{lepto}.
\vglue 2cm

{\elevenbf \noindent Acknowledgments \hfil}
\vglue 0.4cm
We thank S.Rindani and R.M.Godbole for valuable discussions.
A.S.J. wants to thank A. Masiero for helpful discussions.
M.N. wishes to thank the Alexander von Humboldt foundation for
financial support under the Feodor-Lynen Fellowship program.


\begin{thebibliography}{99}
\bibitem{note1}
For different $\mu_i$'s we would have a new term in the
potential with off-diagonal terms of the form $\mu_{ij}(L^{\dagger}_iL_j)$.
\bibitem{note2}
We note here that eqs. (15) and (16) are scale dependent. Furthermore
adding a one loop correction to the tree level potential can
convert the equality to an inequality hereby relaxing the fine
tuning which would otherwise follow from (15) or (16).
\end{thebibliography}
\end{document}